# Reconsideration of photonic tunneling through undersized waveguides

Zhi-Yong Wang[*]

[*]E-mail: zywang@uestc.edu.cn

*School of Optoelectronic Science and Engineering, University of Electronic Science and Technology of China, Chengdu 610054, CHINA*

All the previous studies on photonic tunneling are just based on a simple and directly analogy with a one-dimensional quantum-mechanical tunneling, without taking into account the horizontal structure of electromagnetic waves along the waveguide, such that they are oversimplified and incomplete. Here we present a more serious deliberation on photonic tunneling through cut-off waveguides, and obtain a strictly theoretical model with some new results.

## I. INTRODUCTION

Optical analog for quantum-mechanical barrier penetration, i.e., photonic tunneling, has been studied by many researchers [1-3] since electronic tunneling in solid state physics was evidenced in the 1950s. In particular, the question of how long it takes a particle to tunnel through a potential barrier has been investigated extensively since quantum mechanics was founded [4-11], and such tunneling time problem has experienced a novel stimulus by the results of analogous experiments with evanescent electromagnetic wave packets. By means of photonic tunneling, many experimenters have reported measuring superluminal tunneling velocities, which is in agreement with the fact that many tunneling time definitions seem to predict superluminal tunneling velocities (such as the Hartman effect), and causes massive debate [10-21]. On the whole, as an electromagnetic analog of the quantum-mechanical tunnel effect, photonic tunneling play a very important role for the issue of tunneling time, and is related to surface plasmon polaritons, i.e., the key concept in nanophotonics.

However, all the previous theoretical models of photonic tunneling have been established just by means of a simple and directly analogy with a one-dimensional quantum-mechanical tunneling, without taking into account the horizontal components of electromagnetic waves (i.e., , the components perpendicular to the direction of propagation), and ignoring the fact that there are different boundary conditions for different components. As a result, the original works on photonic tunneling are oversimplified and incomplete. In this paper, we present a more serious deliberation on photonic tunneling through cut-off waveguides, and obtain a strictly theoretical model with some new results.

The paper is organized as follows. In Section II, we introduce a rectangular waveguide with discontinuities and review the electromagnetic field inside the waveguide, including its special continuity conditions and orthonormality. This is especially important as it will make clear what the transmission part of a given mode consists of. In the most general way, we study the reflection and transmission coefficients for the TE and TM modes, respectively. In Section III, two specific examples are presented, where the incident electromagnetic waves are taken as $TE_{10}$ and $TM_{11}$ modes, respectively. Appendices A and B are helpful for the text.

## II. ELECTRMAGNETIC WAVES PROPAGATING ALONG A RECTANGULAR WAVEGUIDE WITH DISCONTINUITIES

To study the reflection and transmission of electromagnetic waves propagating along a discontinuous waveguide from $z<0$ to $z>L$, through the region of $0<z<L$, let us assume that the waveguide has perfectly conducting walls and is placed along the z-axis of rectangular coordinates $(x,y,z)$, and it is filled with lossless insulator media whose permittivity and permeability are real constants. Specifically, in the regions of $z<0$ and $z>L$, the waveguide has cross sectional dimensions $a\times b$ ($a\geq b$, $x\in[0,a]$ and $y \in [0,b]$ inside the waveguide), and its filling medium possesses the relative permittivity $\varepsilon_r$ and the relative permeability $\mu_r$, respectively; while, as $0 < z < L$, the cross sectional dimensions becomes $s \times d$ ($s \geq d$, $x \in [0,s]$ and $y \in [0,d]$ inside the waveguide), and the relative permittivity and permeability of the filling media become $\varepsilon_r'$ and $\mu_r'$, respectively. Moreover, let us suppose that $a > s$ and $b \geq d$. Let $\omega$ and $\boldsymbol{k} = (k_x, k_y, k_z)$ denote the frequency and the wave-number vector of electromagnetic waves propagating along the waveguide, respectively, where

1) As $z < 0$ or $z > L$, let $k_{\perp mn}^2 = k_x^2 + k_y^2$, one has $|k_x| = m\pi/a$, $|k_y| = n\pi/b$, $k_z = h_{mn} = \pm\sqrt{\varepsilon_r \mu_r \omega^2/c^2 - k_{\perp mn}^2}$, where, for the TE$_{mn}$ mode one has $m, n = 0, 1, 2, ...$ except $m = n = 0$, while for the TM$_{mn}$ mode one has $m, n = 1, 2, ...$;

2) As $0 \leq z \leq L$, let $k_{\perp pq}'^2 = k_x^2 + k_y^2$, one has $|k_x| = p\pi/s$, $|k_y| = q\pi/d$, $k_z = h_{pq}' = \pm\sqrt{\varepsilon_r' \mu_r' \omega^2/c^2 - k_{\perp pq}'^2}$, where, for the TE$_{pq}$ mode one has $p, q = 0, 1, 2, ...$ except $p = q = 0$, while for the TM$_{pq}$ mode one has $p, q = 1, 2, ...$.

As $z < 0$ or $z > L$, according to electromagnetic waveguide theory, the TE$_{mn}$ and TM$_{mn}$ modes can be expressed as, respectively

$$\text{TE}_{mn}: \begin{cases} E_{x,mn} = -\mathrm{i}\dfrac{\omega\mu_0\mu_r}{k_{\perp mn}^2}\dfrac{n\pi}{b}H_{0,mn}\cos(\dfrac{m\pi}{a}x)\sin(\dfrac{n\pi}{b}y)\exp[-\mathrm{i}(\omega t - h_{mn}z)] \\ E_{y,mn} = \mathrm{i}\dfrac{\omega\mu_0\mu_r}{k_{\perp mn}^2}\dfrac{m\pi}{a}H_{0,mn}\sin(\dfrac{m\pi}{a}x)\cos(\dfrac{n\pi}{b}y)\exp[-\mathrm{i}(\omega t - h_{mn}z)] \\ E_{z,mn} = 0 \\ H_{x,mn} = -\mathrm{i}\dfrac{h_{mn}}{k_{\perp mn}^2}\dfrac{m\pi}{a}H_{0,mn}\sin(\dfrac{m\pi}{a}x)\cos(\dfrac{n\pi}{b}y)\exp[-\mathrm{i}(\omega t - h_{mn}z)] \\ H_{y,mn} = -\mathrm{i}\dfrac{h_{mn}}{k_{\perp mn}^2}\dfrac{n\pi}{b}H_{0,mn}\cos(\dfrac{m\pi}{a}x)\sin(\dfrac{n\pi}{b}y)\exp[-\mathrm{i}(\omega t - h_{mn}z)] \\ H_{z,mn} = H_{0,mn}\cos(\dfrac{m\pi}{a}x)\cos(\dfrac{n\pi}{b}y)\exp[-\mathrm{i}(\omega t - h_{mn}z)] \end{cases} \quad (1)$$

$$\text{TM}_{mn}: \begin{cases} E_{x,mn} = \mathrm{i}\dfrac{h_{mn}}{k_{\perp mn}^2}\dfrac{m\pi}{a}E_{0,mn}\cos(\dfrac{m\pi}{a}x)\sin(\dfrac{n\pi}{b}y)\exp[-\mathrm{i}(\omega t - h_{mn}z)] \\[4pt] E_{y,mn} = \mathrm{i}\dfrac{h_{mn}}{k_{\perp mn}^2}\dfrac{n\pi}{b}E_{0,mn}\sin(\dfrac{m\pi}{a}x)\cos(\dfrac{n\pi}{b}y)\exp[-\mathrm{i}(\omega t - h_{mn}z)] \\[4pt] E_{z,mn} = E_{0,mn}\sin(\dfrac{m\pi}{a}x)\sin(\dfrac{n\pi}{b}y)\exp[-\mathrm{i}(\omega t - h_{mn}z)] \\[4pt] H_{x,mn} = -\mathrm{i}\dfrac{\omega\varepsilon_0\varepsilon_r}{k_{\perp mn}^2}\dfrac{n\pi}{b}E_{0,mn}\sin(\dfrac{m\pi}{a}x)\cos(\dfrac{n\pi}{b}y)\exp[-\mathrm{i}(\omega t - h_{mn}z)] \\[4pt] H_{y,mn} = \mathrm{i}\dfrac{\omega\varepsilon_0\varepsilon_r}{k_{\perp mn}^2}\dfrac{m\pi}{a}E_{0,mn}\cos(\dfrac{m\pi}{a}x)\sin(\dfrac{n\pi}{b}y)\exp[-\mathrm{i}(\omega t - h_{mn}z)] \\[4pt] H_{z,mn} = 0 \end{cases} \quad (2)$$

Where $H_{0,mn}$ and $E_{0,mn}$ are two constants. As $0 \leq z \leq L$, the expressions of the $\text{TE}_{mn}$ and $\text{TM}_{mn}$ modes are similar to Eqs. (1) and (2). In terms of Eqs. (1) and (2), one can study the electromagnetic field inside the waveguide as a whole, and as the solution of Maxwell equations, it must satisfy all the constraints between field components. Unfortunately, people have usually focused on just a single component of the electromagnetic field, and study all of its possible solutions without taking into account the constraints between it and the other components (e.g., in non-Hermitian photonics). However, the true physical solutions of the single component must simultaneously satisfy five other equations, and some of the so-called miraculous results are actually spurious ones.

Now, let us consider that an electromagnetic wave of a given $\text{TE}_{mn}$ (or $\text{TM}_{mn}$) eigenmode propagates along the waveguide from the region of $z < 0$ to that of $0 < z < L$. In general, a part of the incident electromagnetic wave will be reflected back to the region of $z < 0$, while the other part will transmit through the region of $0 < z < L$ and enter into the region of $z > L$. Taking into consideration (see **Appendices A** and **B**, $i = 1, 2, 3, 4$)

$$\int_0^a\int_0^b g_{mn}^{(i)}(x,y)g_{m'n'}^{(i)}(x,y)\mathrm{d}x\mathrm{d}y = \frac{ab}{4}\delta_{mm'}\delta_{nn'}, \quad \int_0^b\int_0^a \psi_{\text{TE}_{mn}}^{\dagger}\psi_{\text{TM}_{pq}}\mathrm{d}x\mathrm{d}y = 0,$$

that is, the TE and TM modes are orthogonal. As a result, when the transmission part of the given $TE_{mn}$ (or $TM_{mn}$) wave propagates into the region of $0 < z < L$, it should be expressed as the linear superposition of the $TE_{pq}$ (or $TM_{pq}$) eigenmodes in the region of $0 < z < L$ (for the moment $x \in [0, s]$ and $y \in [0, d]$), while the transition probability between the TE and TM modes can be neglected.

However, according to the relationships between electromagnetic field components on interfaces, only the components of $E_x$, $E_y$ and $H_z$ are continuous, and then we will just discuss the boundary conditions of $E_x$, $E_y$ and $H_z$, from which we will obtain the reflection and transmission coefficients, as well as the dispersion relations, etc.. Then for our purpose, let the superscripts (+) and (−) denote that the electromagnetic waves propagate along the positive and negative directions of the waveguide, respectively. The three continuous components of $E_x$, $E_y$ and $H_z$ in the $TE_{mn}$ mode can together be expressed as

$$\begin{pmatrix} E_{x,mn}^{(\pm)} \\ E_{y,mn}^{(\pm)} \\ H_{z,mn}^{(\pm)} \end{pmatrix} = H_{0,mn}^{(\pm)} f_{TE}(x,y) \exp[-i(\omega t \mp h_{mn} z)], \qquad (3)$$

where, it follows from Eq. (1) that, for $m, n = 0, 1, 2, ...$ but for $m = n = 0$,

$$f_{TE}(x,y) = \begin{pmatrix} -i\dfrac{\omega \mu_0 \mu_r}{k_{\perp mn}^2} \dfrac{n\pi}{b} \cos(\dfrac{m\pi}{a} x) \sin(\dfrac{n\pi}{b} y) \\ i\dfrac{\omega \mu_0 \mu_r}{k_{\perp mn}^2} \dfrac{m\pi}{a} \sin(\dfrac{m\pi}{a} x) \cos(\dfrac{n\pi}{b} y) \\ \cos(\dfrac{m\pi}{a} x) \cos(\dfrac{n\pi}{b} y) \end{pmatrix}. \qquad (4)$$

Likewise, the two continuous components of $E_x$ and $E_y$ in the $TM_{mn}$ mode can together be expressed as (taking no account of $H_z = 0$)

$$\begin{pmatrix} E_{x,mn}^{(\pm)} \\ E_{y,mn}^{(\pm)} \end{pmatrix} = E_{0,mn}^{(\pm)} f_{TM}(x,y) \exp[-i(\omega t \mp h_{mn} z)], \qquad (5)$$

where, it follows from Eq. (2) that, for $m, n = 1, 2, 3, \ldots$,

$$f_{\text{TM}}(x, y) = \begin{pmatrix} \mathrm{i} \dfrac{h_{mn}}{k_{\perp mn}^2} \dfrac{m\pi}{a} \cos(\dfrac{m\pi}{a} x) \sin(\dfrac{n\pi}{b} y) \\ \mathrm{i} \dfrac{h_{mn}}{k_{\perp mn}^2} \dfrac{n\pi}{b} \sin(\dfrac{m\pi}{a} x) \cos(\dfrac{n\pi}{b} y) \end{pmatrix}. \tag{6}$$

In the following we will study the reflection and transmission coefficients for the TE and TM modes, respectively.

**A. The incident electromagnetic wave as the TE mode**

For the $\text{TE}_{mn}$ mode, in terms of its three continuous components $E_x$, $E_y$ and $H_z$, as discussed above one can express the electromagnetic waves propagating along the waveguide from the region of $z < 0$ to that of $z > L$ as follows:

$$\begin{cases} \varphi_1(x, y, z) = [A_{1mn} \exp(\mathrm{i} h_{mn} z) + A_{2mn} \exp(-\mathrm{i} h_{mn} z)] f_{\text{TE}_{mn}}(x, y), \; z < 0 \\ \varphi_2(x, y, z) = \sum_{pq} [B_{1pq} \exp(\mathrm{i} h'_{pq} z) + B_{2pq} \exp(-\mathrm{i} h'_{pq} z)] f'_{\text{TE}_{pq}}(x, y), \; 0 < z < L, \\ \varphi_3(x, y, z) = C_{mn} \exp(\mathrm{i} h_{mn} z) f_{\text{TE}_{mn}}(x, y), \; z > L \end{cases} \tag{7}$$

where $A_{1mn}$, $A_{2mn}$, $B_{1pq}$, $B_{2pq}$ and $C_{mn}$ are constants, $m, n, p, q = 0, 1, 2, \ldots$, except $m = n = 0$ and $p = q = 0$, $f_{\text{TE}_{mn}}(x, y)$ is given by Eq. (4), and

$$f'_{\text{TE}_{pq}} = \begin{pmatrix} -\mathrm{i} \dfrac{\omega \mu_0 \mu'_r}{k'^2_{\perp pq}} \dfrac{q\pi}{d} \cos(\dfrac{p\pi}{s} x) \sin(\dfrac{q\pi}{d} y) \\ \mathrm{i} \dfrac{\omega \mu_0 \mu'_r}{k'^2_{\perp pq}} \dfrac{p\pi}{s} \sin(\dfrac{p\pi}{s} x) \cos(\dfrac{q\pi}{d} y) \\ \cos(\dfrac{p\pi}{s} x) \cos(\dfrac{q\pi}{d} y) \end{pmatrix}. \tag{8}$$

Here we have omitted the factor of $\exp(-\mathrm{i}\omega t)$, and the constants $H_{0,mn}^{(+)}$ and $H_{0,mn}^{(-)}$ are absorbed by the coefficients $A_{1mn}$ and $A_{2mn}$, respectively, and so on. Moreover,

$$k_{\perp mn}^2 = (m\pi/a)^2 + (n\pi/b)^2, \; h_{mn} = \sqrt{\omega^2/u^2 - k_{\perp mn}^2}, \; u = 1 \Big/ \sqrt{\varepsilon_0 \mu_0 \varepsilon_r \mu_r}, \tag{9}$$

$$k'^2_{\perp pq} = (p\pi/s)^2 + (q\pi/d)^2, \; h'_{pq} = \sqrt{\omega^2/u'^2 - k'^2_{\perp pq}}, \; u' = 1 \Big/ \sqrt{\varepsilon_0 \mu_0 \varepsilon'_r \mu'_r}. \tag{10}$$

The continuity conditions are

$$\begin{cases} \varphi_1\big|_{z=0} = \varphi_2\big|_{z=0}, & \dfrac{d\varphi_1}{dz}\bigg|_{z=0} = \dfrac{d\varphi_2}{dz}\bigg|_{z=0} \\ \varphi_3\big|_{z=L} = \varphi_2\big|_{z=L}, & \dfrac{d\varphi_3}{dz}\bigg|_{z=L} = \dfrac{d\varphi_2}{dz}\bigg|_{z=L} \end{cases}. \tag{11}$$

Because of $a \geq s$ and $b \geq d$, at the entrance and exit of the intermediate waveguide (i.e., at $z = 0$ and $z = L$, respectively), the conditions of continuity are valid for $x \in [0, s]$ and $y \in [0, d]$, and then we just consider the incident wave that appears within the cross section of $x \in [0, s]$ and $y \in [0, d]$. It follows from Eq. (11) that

$$\begin{cases} (A_{1mn} + A_{2mn}) f_{TE_{mn}} = \sum_{pq}(B_{1pq} + B_{2pq}) f'_{TE_{pq}} \\ (h_{mn}A_{1mn} - h_{mn}A_{2mn}) f_{TE_{mn}} = \sum_{pq}(h'_{pq}B_{1pq} - h'_{pq}B_{2pq}) f'_{TE_{pq}} \\ C_{mn}\exp(ih_{mn}L) f_{TE_{mn}} = \sum_{pq}[B_{1pq}\exp(ih'_{pq}L) + B_{2pq}\exp(-ih'_{pq}L)] f'_{TE_{pq}} \\ h_{mn}C_{mn}\exp(ih_{mn}L) f_{TE_{mn}} = \sum_{pq}[h'_{pq}B_{1pq}\exp(ih'_{pq}L) - h'_{pq}B_{2pq}\exp(-ih'_{pq}L)] f'_{TE_{pq}} \end{cases}. \tag{12}$$

It is easy to prove the following formulae

$$\begin{cases} \int_0^s \cos(\dfrac{p\pi}{s}x)\cos(\dfrac{p'\pi}{s}x)dx = \int_0^s \sin(\dfrac{p\pi}{s}x)\sin(\dfrac{p'\pi}{s}x)dx = \dfrac{s}{2}\delta_{pp'} \\ \int_0^d \cos(\dfrac{q\pi}{d}y)\cos(\dfrac{q'\pi}{d}y)dy = \int_0^d \sin(\dfrac{q\pi}{d}y)\sin(\dfrac{q'\pi}{d}y)dy = \dfrac{d}{2}\delta_{qq'} \end{cases}, \tag{13}$$

$$\begin{cases} \int_0^s \cos(\dfrac{m\pi}{a}x)\cos(\dfrac{p'\pi}{s}x)dx = \dfrac{mas^2(-1)^{p'+1}}{\pi(p'a+ms)(p'a-ms)}\sin\dfrac{ms\pi}{a} \\ \int_0^s \sin(\dfrac{m\pi}{a}x)\sin(\dfrac{p'\pi}{s}x)dx = \dfrac{p'a^2 s(-1)^{p'+1}}{\pi(p'a+ms)(p'a-ms)}\sin\dfrac{ms\pi}{a} \\ \int_0^d \cos(\dfrac{n\pi}{b}y)\cos(\dfrac{q'\pi}{d}y)dy = \dfrac{nbd^2(-1)^{q'+1}}{\pi(q'b+nd)(q'b-nd)}\sin\dfrac{nd\pi}{b} \\ \int_0^d \sin(\dfrac{n\pi}{b}y)\sin(\dfrac{q'\pi}{d}y)dy = \dfrac{q'b^2 d(-1)^{q'+1}}{\pi(q'b+nd)(q'b-nd)}\sin\dfrac{nd\pi}{b} \end{cases}, \tag{14}$$

Applying Eqs. (12)-(14), one can obtain, *under the condition of* $k^2_{\perp mn}\mu'_r = k'^2_{\perp pq}\mu_r$,

$$\begin{cases} (A_{1mn} + A_{2mn})\Gamma_{mnpq} = B_{1pq} + B_{2pq} \\ (h_{mn}A_{1mn} - h_{mn}A_{2mn})\Gamma_{mnpq} = h'_{pq}B_{1pq} - h'_{pq}B_{2pq} \\ C_{mn}\exp(ih_{mn}L)\Gamma_{mnpq} = B_{1pq}\exp(ih'_{pq}L) + B_{2pq}\exp(-ih'_{pq}L) \\ h_{mn}C_{mn}\exp(ih_{mn}L)\Gamma_{mnpq} = h'_{pq}B_{1pq}\exp(ih'_{pq}L) - h'_{pq}B_{2pq}\exp(-ih'_{pq}L) \end{cases}, \quad (15)$$

where

$$\Gamma_{mnpq} = \frac{4mnabsd(-1)^{p+q}}{\pi^2(p^2a^2 - m^2s^2)(q^2b^2 - n^2d^2)}\sin\frac{ms\pi}{a}\sin\frac{nd\pi}{b}. \quad (16)$$

Let

$$A'_{1mn} = \Gamma_{mnpq}A_{1mn}, \quad A'_{2mn} = \Gamma_{mnpq}A_{2mn}, \quad C'_{mn} = \Gamma_{mnpq}C_{mn}. \quad (17)$$

Eq. (15) can be rewritten as

$$\begin{cases} A'_{1mn} + A'_{2mn} = B_{1pq} + B_{2pq} \\ h_{mn}A'_{1mn} - h_{mn}A'_{2mn} = h'_{pq}B_{1pq} - h'_{pq}B_{2pq} \\ C'_{mn}\exp(ih_{mn}L) = B_{1pq}\exp(ih'_{pq}L) + B_{2pq}\exp(-ih'_{pq}L) \\ h_{mn}C'_{mn}\exp(ih_{mn}L) = h'_{pq}B_{1pq}\exp(ih'_{pq}L) - h'_{pq}B_{2pq}\exp(-ih'_{pq}L) \end{cases} \quad \text{with } \frac{k^2_{\perp mn}\mu'_r}{k'^2_{\perp pq}\mu_r} = 1. \quad (18)$$

It follows from Eq. (18) that

$$\begin{cases} C'_{mn} = \dfrac{4h_{mn}h'_{pq}\exp(-ih_{mn}L)}{(h_{mn}+h'_{pq})^2\exp(-ih'_{pq}L) - (h_{mn}-h'_{pq})^2\exp(ih'_{pq}L)}A'_{1mn} \\ A'_{2mn} = \dfrac{2i(h^2_{mn} - h'^2_{pq})\sin h'_{pq}L}{(h_{mn}-h'_{pq})^2\exp(ih'_{pq}L) - (h_{mn}+h'_{pq})^2\exp(-ih'_{pq}L)}A'_{1mn} \end{cases}, \quad (19)$$

$$\begin{cases} B_{1pq} = \dfrac{2h_{mn}(h_{mn}+h'_{pq})\exp(-ih'_{pq}L)}{(h_{mn}+h'_{pq})^2\exp(-ih'_{pq}L) - (h_{mn}-h'_{pq})^2\exp(ih'_{pq}L)}A'_{1mn} \\ B_{2pq} = \dfrac{-2h_{mn}(h_{mn}-h'_{pq})\exp(ih'_{pq}L)}{(h_{mn}+h'_{pq})^2\exp(-ih'_{pq}L) - (h_{mn}-h'_{pq})^2\exp(ih'_{pq}L)}A'_{1mn} \end{cases}. \quad (20)$$

Substituting Eq. (17) into Eqs. (19) and (20), one has, respectively

$$\begin{cases} C_{mn} = \dfrac{4h_{mn}h'_{pq}\exp(-ih_{mn}L)}{(h_{mn}+h'_{pq})^2\exp(-ih'_{pq}L) - (h_{mn}-h'_{pq})^2\exp(ih'_{pq}L)}A_{1mn} \\ A_{2mn} = \dfrac{2i(h^2_{mn} - h'^2_{pq})\sin h'_{pq}L}{(h_{mn}-h'_{pq})^2\exp(ih'_{pq}L) - (h_{mn}+h'_{pq})^2\exp(-ih'_{pq}L)}A_{1mn} \end{cases}. \quad (21)$$

$$\begin{cases} B_{1pq} = \dfrac{2h_{mn}(h_{mn}+h'_{pq})\exp(-ih'_{pq}L)}{(h_{mn}+h'_{pq})^2\exp(-ih'_{pq}L)-(h_{mn}-h'_{pq})^2\exp(ih'_{pq}L)}\Gamma_{mnpq}A_{1mn} \\ B_{2pq} = \dfrac{-2h_{mn}(h_{mn}-h'_{pq})\exp(ih'_{pq}L)}{(h_{mn}+h'_{pq})^2\exp(-ih'_{pq}L)-(h_{mn}-h'_{pq})^2\exp(ih'_{pq}L)}\Gamma_{mnpq}A_{1mn} \end{cases}. \quad (22)$$

Using Eq. (21) one can obtain

$$\begin{cases} T_{pq} = \dfrac{|C_{mn}|^2}{|A_{1mn}|^2} = \dfrac{4h_{mn}^2 h'^2_{pq}}{(h_{mn}^2-h'^2_{pq})^2\sin^2 h'_{pq}L + 4h_{mn}^2 h'^2_{pq}} \\ R_{pq} = \dfrac{|A_{2mn}|^2}{|A_{1mn}|^2} = \dfrac{(h_{mn}^2-h'^2_{pq})^2\sin^2 h'_{pq}L}{(h_{mn}^2-h'^2_{pq})^2\sin^2 h'_{pq}L + 4h_{mn}^2 h'^2_{pq}} \end{cases}. \quad (23)$$

Eqs. (23) is valid for $\varepsilon'_r\mu'_r\omega^2/c^2 > (p\pi/s)^2+(q\pi/d)^2$, such that $h'_{pq}$ is a real number. On the other hand, if $\varepsilon'_r\mu'_r\omega^2/c^2 < (p\pi/s)^2+(q\pi/d)^2$, let $h'_{pq} = i\kappa_{pq}$, where $\kappa_{pq}$ is a real number and

$$\kappa_{pq} = \sqrt{(\dfrac{p\pi}{s})^2+(\dfrac{q\pi}{d})^2 - \varepsilon'_r\mu'_r\dfrac{\omega^2}{c^2}}. \quad (24)$$

For the moment one has

$$\begin{cases} T_{pq} = \dfrac{|C_{mn}|^2}{|A_{1mn}|^2} = \dfrac{4h_{mn}^2\kappa_{pq}^2}{(h_{mn}^2+\kappa_{pq}^2)^2\sinh^2\kappa_{pq}L + 4h_{mn}^2\kappa_{pq}^2} \\ R_{pq} = \dfrac{|A_{2mn}|^2}{|A_{1mn}|^2} = \dfrac{(h_{mn}^2+\kappa_{pq}^2)^2\sinh^2\kappa_{pq}L}{(h_{mn}^2+\kappa_{pq}^2)^2\sinh^2\kappa_{pq}L + 4h_{mn}^2\kappa_{pq}^2} \end{cases}. \quad (25)$$

Note that Eqs. (23) and (25) are valid only under the condition of $k^2_{\perp mn}\mu'_r = k'^2_{\perp pq}\mu_r$.

Now, let us explain the physical meanings of $R_{pq}$ and $T_{pq}$. The z-component of the time-averaged Poynting vector along the waveguide is

$$S_z = \dfrac{1}{2}\text{Re}(\boldsymbol{E}\times\boldsymbol{H}^*)_z = \dfrac{1}{2}\text{Re}(E_x H_y^* - E_y H_x^*). \quad (26)$$

Eq.(26) implies that $S_z \propto |A_{1mn}|^2, |A_{2mn}|^2, |C_{mn}|^2$, such that the parameters $R_{pq}$ and $T_{pq}$, given by Eq.(23) or Eq. (25), respectively describe the reflectivity and transmissivity of the incident wave (when it appears in the cross section of $x\in[0,s]$ and $y\in[0,d]$), via the $\mathrm{TE}_{pq}$ mode excited inside the intermediate waveguide within $0 < z < L$.

It is important to note that, the cross-sectional dimension $s \times d$ of the intermediate waveguide is less than $a \times b$ (i.e., the cross-sectional dimension of the waveguides for $z < 0$ and $z > L$), and then there is also another type of reflection on the wall where $x \in [s, a]$ and $y \in [d, a]$. However, this reflection on the transverse dividing wall of $z = 0$ with $x \in [s, a]$ and $y \in [d, a]$, has nothing to do with the coefficients $R_{pq}$ and $T_{pq}$, and it is irrelevant to our subject. In other words, at the entrance and exit of the intermediate waveguide (i.e., at $z=0$ and $z=L$, respectively), the conditions of continuity are valid for $x \in [0, s]$ and $y \in [0, d]$, such that we only consider the incident wave that appears in the cross section of $x \in [0, s]$ and $y \in [0, d]$.

**B. The incident electromagnetic wave as the TM mode**

Similarly, for the $\mathrm{TM}_{mn}$ mode, in terms of its two continuous components $E_x$ and $E_y$ one can express the electromagnetic waves propagating along the waveguide from the region of $z < 0$ to that of $z > L$ as follows:

$$\begin{cases} \varphi_1(x,y,z) = [A_{1mn}\exp(\mathrm{i}h_{mn}z) + A_{2mn}\exp(-\mathrm{i}h_{mn}z)]f_{\mathrm{TM}_{mn}}(x,y),\ z<0 \\ \varphi_2(x,y,z) = \sum_{pq}[B_{1pq}\exp(\mathrm{i}h'_{pq}z) + B_{2pq}\exp(-\mathrm{i}h'_{pq}z)]f'_{\mathrm{TM}_{pq}}(x,y),\ 0<z<L, \\ \varphi_3(x,y,z) = C_{mn}\exp(\mathrm{i}h_{mn}z)f_{\mathrm{TM}_{mn}}(x,y),\ z>L \end{cases} \quad (27)$$

where $A_{1mn}$, $A_{2mn}$, $B_{1pq}$, $B_{2pq}$ and $C_{mn}$ are constants, $m,n,p,q = 1,2,...$, $f_{\mathrm{TM}_{mn}}(x,y)$ is given by Eq. (6), and

$$f'_{\mathrm{TM}_{pq}} = \begin{pmatrix} \mathrm{i}\dfrac{h'_{pq}}{k'^{2}_{\perp pq}}\dfrac{p\pi}{s}\cos(\dfrac{p\pi}{s}x)\sin(\dfrac{q\pi}{d}y) \\ \mathrm{i}\dfrac{h'_{pq}}{k'^{2}_{\perp pq}}\dfrac{q\pi}{d}\sin(\dfrac{p\pi}{s}x)\cos(\dfrac{q\pi}{d}y) \end{pmatrix}. \quad (28)$$

Likewise, we have omitted the factor of $\exp(-\mathrm{i}\omega t)$, and the constants $E^{(+)}_{0,mn}$ and $E^{(-)}_{0,mn}$

are absorbed by the coefficients $A_{1mn}$ and $A_{2mn}$, respectively, and so on. For the moment the continuity conditions are also given by Eq. (11), and ($m, n, p, q = 1, 2, 3, ...$)

$$\begin{cases} k_{\perp mn}^2 = (m\pi/a)^2 + (n\pi/b)^2, h_{mn} = \sqrt{\varepsilon_r \mu_r \omega^2/c^2 - (m\pi/a)^2 - (n\pi/b)^2} \\ k_{\perp pq}'^2 = (p\pi/s)^2 + (q\pi/d)^2, h_{pq}' = \sqrt{\varepsilon_r' \mu_r' \omega^2/c^2 - (p\pi/s)^2 - (q\pi/d)^2} \end{cases}. \quad (29)$$

Substituting Eq. (27) into Eq. (11), one has

$$\begin{cases} (A_{1mn} + A_{2mn}) f_{TM_{mn}} = \sum_{pq} (B_{1pq} + B_{2pq}) f_{TM_{pq}}' \\ (h_{mn} A_{1mn} - h_{mn} A_{2mn}) f_{TM_{mn}} = \sum_{pq} (h_{pq}' B_{1pq} - h_{pq}' B_{2pq}) f_{TM_{pq}}' \\ C_{mn} \exp(ih_{mn} L) f_{TM_{mn}} = \sum_{pq} [B_{1pq} \exp(ih_{pq}' L) + B_{2pq} \exp(-ih_{pq}' L)] f_{TM_{pq}}' \\ h_{mn} C_{mn} \exp(ih_{mn} L) f_{TM_{mn}} = \sum_{pq} [h_{pq}' B_{1pq} \exp(ih_{pq}' L) - h_{pq}' B_{2pq} \exp(-ih_{pq}' L)] f_{TM_{pq}}' \end{cases}. \quad (30)$$

Applying Eqs. (13), (14) and (30), one can obtain, *under the condition of* $mbsq = nadp$,

$$\begin{cases} (A_{1mn} + A_{2mn}) \Gamma_{mnpq} = (B_{1pq} + B_{2pq}) \Lambda_{mnpq} \\ (h_{mn} A_{1mn} - h_{mn} A_{2mn}) \Gamma_{mnpq} = (h_{pq}' B_{1pq} - h_{pq}' B_{2pq}) \Lambda_{mnpq} \\ C_{mn} \exp(ih_{mn} L) \Gamma_{mnpq} = [B_{1pq} \exp(ih_{pq}' L) + B_{2pq} \exp(-ih_{pq}' L)] \Lambda_{mnpq} \\ h_{mn} C_{mn} \exp(ih_{mn} L) \Gamma_{mnpq} = [h_{pq}' B_{1pq} \exp(ih_{pq}' L) - h_{pq}' B_{2pq} \exp(-ih_{pq}' L)] \Lambda_{mnpq} \end{cases}, \quad (31)$$

where $\Gamma_{mnpq}$ is given by Eq. (16), and

$$\Lambda_{mnpq} = \frac{k_{\perp mn}^2 h_{pq}'}{k_{\perp pq}'^2 h_{mn}}. \quad (32)$$

Similarly, let

$$A_{imn}' = \Gamma_{mnpq} A_{imn}, \quad B_{ipq}' = \Lambda_{mnpq} B_{ipq}, \quad C_{mn}' = \Gamma_{mnpq} C_{mn}, \quad i = 1, 2. \quad (33)$$

Eq. (31) can be rewritten as

$$\begin{cases} A_{1mn}' + A_{2mn}' = B_{1pq}' + B_{2pq}' \\ h_{mn}(A_{1mn}' - A_{2mn}') = h_{pq}'(B_{1pq}' - B_{2pq}') \\ C_{mn}' \exp(ih_{mn} L) = B_{1pq}' \exp(ih_{pq}' L) + B_{2pq}' \exp(-ih_{pq}' L) \\ h_{mn} C_{mn}' \exp(ih_{mn} L) = h_{pq}'[B_{1pq}' \exp(ih_{pq}' L) - B_{2pq}' \exp(-ih_{pq}' L)] \end{cases} \quad \text{with} \quad \frac{mbsq}{nadp} = 1. \quad (34)$$

Applying Eq. (34) one has

1) As $h'_{pq} = \sqrt{\varepsilon'_r \mu'_r \omega^2/c^2 - (p\pi/s)^2 - (q\pi/d)^2}$ is a real number, the reflectivity of $R_{pq}$ and the transmissivity of $T_{pq}$ are, respectively,

$$\begin{cases} T_{pq} = \dfrac{|C'_{mn}|^2}{|A'_{1mn}|^2} = \dfrac{|C_{mn}|^2}{|A_{1mn}|^2} = \dfrac{4h_{mn}^2 h'^2_{pq}}{(h_{mn}^2 - h'^2_{pq})^2 \sin^2 h'_{pq} L + 4h_{mn}^2 h'^2_{pq}} \\ R_{pq} = \dfrac{|A'_{2mn}|^2}{|A'_{1mn}|^2} = \dfrac{|A_{2mn}|^2}{|A_{1mn}|^2} = \dfrac{(h_{mn}^2 - h'^2_{pq})^2 \sin^2 h'_{pq} L}{(h_{mn}^2 - h'^2_{pq})^2 \sin^2 h'_{pq} L + 4h_{mn}^2 h'^2_{pq}} \end{cases}. \quad (35)$$

2) As $\kappa_{pq} = \sqrt{(p\pi/s)^2 + (q\pi/d)^2 - \varepsilon'_r \mu'_r \omega^2/c^2}$ is a real number, the reflectivity of $R_{pq}$ and the transmissivity of $T_{pq}$ are, respectively,

$$\begin{cases} T_{pq} = \dfrac{|C'_{mn}|^2}{|A'_{1mn}|^2} = \dfrac{|C_{mn}|^2}{|A_{1mn}|^2} = \dfrac{4h_{mn}^2 \kappa_{pq}^2}{(h_{mn}^2 + \kappa_{pq}^2)^2 \sinh^2 \kappa_{pq} L + 4h_{mn}^2 \kappa_{pq}^2} \\ R_{pq} = \dfrac{|A'_{2mn}|^2}{|A'_{1mn}|^2} = \dfrac{|A_{2mn}|^2}{|A_{1mn}|^2} = \dfrac{(h_{mn}^2 + \kappa_{pq}^2)^2 \sinh^2 \kappa_{pq} L}{(h_{mn}^2 + \kappa_{pq}^2)^2 \sinh^2 \kappa_{pq} L + 4h_{mn}^2 \kappa_{pq}^2} \end{cases}. \quad (36)$$

**III. SOME CONCRETE EXAMPLES**

As two specific examples, let us take the incident electromagnetic waves, propagating along the waveguide from the region of $z < 0$ to that of $0 < z < L$, as the $TE_{10}$ and $TM_{11}$ modes, respectively. Because of $R_{pq} + T_{pq} = 1$, we will just discuss the transmission coefficient $T_{pq}$.

1) The incident electromagnetic wave as the TE$_{10}$ mode

In this case one has $m = 1$ and $n = 0$. For convenience let $\mu_r = \varepsilon'_r = \mu'_r = 1$ and $\varepsilon_r > 1$. The constraint condition of $k_{\perp mn}^2 \mu'_r = k'^2_{\perp pq} \mu_r$ implies that $p = m = 1$, $q = n = 0$ and $a = s$ (for the moment $b = d$ is optional), and then one has

$$k_{\perp 10}^2 = k'^2_{\perp 10} = \pi^2/a^2, \quad h_{10} = \sqrt{\varepsilon_r \omega^2/c^2 - \pi^2/a^2}, \quad h'_{10} = \sqrt{\omega^2/c^2 - \pi^2/a^2}. \quad (37)$$

As $\omega^2/c^2 > \pi^2/a^2$, substituting Eq. (37) into Eq. (35) one can obtain the transmissivity of $T_{10}$ as follows:

$$T_{10} = \frac{4(\varepsilon_r\omega^2 - c^2\pi^2/a^2)(\omega^2 - c^2\pi^2/a^2)}{(\varepsilon_r - 1)^2 \omega^4 \sin^2 L\sqrt{\omega^2/c^2 - \pi^2/a^2} + 4(\varepsilon_r\omega^2 - c^2\pi^2/a^2)(\omega^2 - c^2\pi^2/a^2)}. \quad (38)$$

As $\omega^2/c^2 < (\pi/a)^2 < \varepsilon_r \omega^2/c^2$, substituting Eq. (37) with $h'_{10} = i\kappa_{10}$ into Eq. (36) one can obtain the transmissivity of $T_{10}$ as follows:

$$T_{10} = \frac{4(\varepsilon_r\omega^2 - c^2\pi^2/a^2)(c^2\pi^2/a^2 - \omega^2)}{(\varepsilon_r - 1)^2 \omega^4 \sinh^2 L\sqrt{\pi^2/a^2 - \omega^2/c^2} + 4(\varepsilon_r\omega^2 - c^2\pi^2/a^2)(c^2\pi^2/a^2 - \omega^2)}. \quad (39)$$

2) The incident electromagnetic wave as the TM$_{11}$ mode

In this case one has $m = n = 1$. To meet the constraint condition of $mbsq = nadp$, as an example, let $a = 2s$ and $b = 2d$, which implies that $q = p$. For convenience let $q = p = 1$, $\varepsilon'_r = \mu'_r = 1$ and $\varepsilon_r, \mu_r \geq 1$, and then one has

$$\begin{cases} k^2_{\perp 11} = (\pi/a)^2 + (\pi/b)^2, \ h_{11} = \sqrt{\varepsilon_r\mu_r\omega^2/c^2 - (\pi/a)^2 - (\pi/b)^2} \\ k'^2_{\perp 11} = (2\pi/a)^2 + (2\pi/b)^2, \ h'_{11} = \sqrt{\omega^2/c^2 - (2\pi/a)^2 - (2\pi/b)^2} \end{cases}. \quad (40)$$

As $\omega^2/c^2 > (2\pi/a)^2 + (2\pi/b)^2$, substituting Eq. (40) into Eq. (35) one can obtain the transmissivity of $T_{11}$ as follows:

$$T_{11} = \frac{\Gamma_{11}}{[\omega^2(\varepsilon_r\mu_r - 1) + 3c^2\pi^2(1/a^2 + 1/b^2)]^2 \sin^2 L\sqrt{\omega^2/c^2 - (2\pi/a)^2 - (2\pi/b)^2} + \Gamma_{11}}. \quad (41)$$

As $\varepsilon_r\mu_r\omega^2/c^2 > (\pi/a)^2 + (\pi/b)^2$ and $\omega^2/c^2 < (2\pi/a)^2 + (2\pi/b)^2$, substituting Eq. (40) with $h'_{11} = i\kappa_{11}$ into Eq. (36) one can obtain the transmissivity of $T_{11}$ as follows:

$$T_{11} = \frac{-\Gamma_{11}}{[\omega^2(\varepsilon_r\mu_r - 1) + 3c^2\pi^2(1/a^2 + 1/b^2)]^2 \sinh^2 L\sqrt{(2\pi/a)^2 + (2\pi/b)^2 - \omega^2/c^2} - \Gamma_{11}}. \quad (42)$$

where

$$\Gamma_{11} = 4(\varepsilon_r\mu_r\omega^2 - c^2\pi^2/a^2 - c^2\pi^2/b^2)(\omega^2 - 4c^2\pi^2/a^2 - 4c^2\pi^2/b^2). \quad (43)$$

## IV. CONCLUSION

In contrast to a simple analogy with a one-dimensional quantum-mechanical tunneling,

we have presented a rigorous research on photonic tunneling by analyzing electromagnetic waves inside a discontinuous rectangular waveguide placed along the Z axis. Comparing with the previous works, we have shown that one cannot neglect the horizontal structure of electromagnetic waves inside the waveguide.

To be specific, in contrary to the fact that wave functions in a one-dimensional quantum-mechanical tunneling are continuous and smooth at boundary points, the electromagnetic waves inside the rectangular waveguide possess six field components (i.e., $\boldsymbol{E}=(E_x,E_y,E_z)$ and $\boldsymbol{H}=(H_x,H_y,H_z)$, say), and only three of them (i.e., $E_x$, $E_y$ and $H_z$) are continuous and smooth at boundary points. On the other hand, all of the six components are associated with each other via Maxwell's equations. The boundary conditions for $E_x$, $E_y$ and $H_z$ must be satisfied simultaneously, such that there are two different constraints on the transmittance formulae for the TE and TM modes, respectively. Fortunately, though the relations between the incidence, reflection and transmission coefficients of the electromagnetic waves inside the waveguide are more complicated than those in quantum-mechanical tunneling, the final expressions for the reflectivity and transmittance are analogous.

Because the six components of the electromagnetic waves inside the waveguide are associated with each other via Maxwell's equations, to analyze the orthonormality and completeness of the TE and TM modes one can describe the electromagnetic field as a whole by means of the $(1,0)\oplus(0,1)$ spinor formalism. In **Appendices A** and **B** we have proven that the TE and TM modes are orthogonal. As a result, when the transmission part of a given $\text{TE}_{mn}$ (or $\text{TM}_{mn}$) mode propagates from the region of $z<0$ into that of $0<z<L$, it should be expressed as the linear superposition of the $\text{TE}_{pq}$ (or $\text{TM}_{pq}$) eigenmodes in the region of $0<z<L$, while the transition probability between the TE and TM modes should be neglected.

As we know, the simplest elementary building blocks for a quantum computer are quantum bits (qubits), i.e., two-level quantum systems [22, 23]. In the next work we will present a new physical implementation of optical qubits via photonic tunneling.

## ACKNOWLEDGMENTS

The research is supported by the Petrol China Innovation Foundation (Grant No: 2017D-5007-0603).

## APPENDIX A

### The $(1,0) \oplus (0,1)$ spinor formalism of the electromagnetic field inside a waveguide

The $(1,0) \oplus (0,1)$ representation of the group SL(2, C) provides a six-component spinor equivalent to the electromagnetic field tensor [24, 25]. By means of the $(1,0) \oplus (0,1)$ description, one can rewrite the Maxwell equations of sourceless electromagnetic fields as the so-called Dirac-like equation, which is beneficial to study some kind of research subjects. For example, by such a spinor representation one can treat the photon field in curved spacetime via spin connection and the tetrad formalism, and study the gravitational spin-orbit coupling of photons more conveniently [25]. In the following, $c = 1/\sqrt{\varepsilon_0 \mu_0}$ denotes the speed of light in vacuum, repeated indices must be summed according to the Einstein rule, and the four-dimensional (4D) space-time metric tensor is chosen as $g_{\mu\nu} = \text{diag}(1,-1,-1,-1)$, $\mu,\nu = 0,1,2,3$.

In a linear medium with the relative permittivity of $\varepsilon_r$ and the relative permeability of $\mu_r$, there are constitutive equations $\boldsymbol{D} = \varepsilon_0 \varepsilon_r \boldsymbol{E}$ and $\boldsymbol{B} = \mu_0 \mu_r \boldsymbol{H}$, and the speed of light in the medium is $u = c/\sqrt{\varepsilon_r \mu_r}$. In terms of the 3×1 matrix forms of $\boldsymbol{E}$ and $\boldsymbol{H}$, the

$(1,0) \oplus (0,1)$ spinor can be expressed as

$$\psi = \frac{1}{\sqrt{2}} \begin{pmatrix} \sqrt{\varepsilon_0 \varepsilon_r} \boldsymbol{E} \\ i\sqrt{\mu_0 \mu_r} \boldsymbol{H} \end{pmatrix}, \quad \boldsymbol{E} = \begin{pmatrix} E_1 \\ E_2 \\ E_3 \end{pmatrix}, \quad \boldsymbol{H} = \begin{pmatrix} H_1 \\ H_2 \\ H_3 \end{pmatrix}. \tag{a1}$$

Let $I_{3\times 3}$ denote the 3×3 identity matrix, and

$$x^\rho = (ut, \boldsymbol{r}) = (ut, x, y, z), \quad \partial_\rho = \partial/\partial x^\rho = (\partial/u\partial t, \nabla), \tag{a2}$$

$$\begin{cases} \zeta^\rho = \partial_\rho \ln\sqrt{\varepsilon_r} = (\zeta^0, \boldsymbol{\zeta}), \ \zeta^0 = \partial \ln\sqrt{\varepsilon_r}/u\partial t, \ \boldsymbol{\zeta} = \nabla \ln\sqrt{\varepsilon_r} \\ \eta^\rho = \partial_\rho \ln\sqrt{\mu_r} = (\eta^0, \boldsymbol{\eta}), \ \eta^0 = \partial \ln\sqrt{\mu_r}/u\partial t, \ \boldsymbol{\eta} = \nabla \ln\sqrt{\mu_r} \end{cases}, \tag{a3}$$

$$A^0 = A_0 = \begin{pmatrix} \zeta^0 & 0 \\ 0 & \eta^0 \end{pmatrix} \otimes I_{3\times 3}, \quad A^i = -A_i = \begin{pmatrix} \zeta^i & 0 \\ 0 & \eta^i \end{pmatrix} \otimes I_{3\times 3}, \quad D_\rho = \partial_\rho + A_\rho. \tag{a4}$$

One can prove that the Maxwell equations can be expressed as the Dirac-like equation

$$i\beta^\rho D_\rho \psi(x) = 0, \tag{a5}$$

where

$$\beta^\rho = (\beta^0, \boldsymbol{\beta}), \text{ with } \beta^0 = \beta_0 = \begin{pmatrix} I_{3\times 3} & 0 \\ 0 & -I_{3\times 3} \end{pmatrix}, \quad \beta^i = -\beta_i = \begin{pmatrix} 0 & \tau_i \\ -\tau_i & 0 \end{pmatrix}, \quad i = 1, 2, 3, \tag{a6}$$

$$\boldsymbol{\tau} = (\tau_1, \tau_2, \tau_3), \quad \tau_1 = \begin{pmatrix} 0 & 0 & 0 \\ 0 & 0 & -i \\ 0 & i & 0 \end{pmatrix}, \quad \tau_2 = \begin{pmatrix} 0 & 0 & i \\ 0 & 0 & 0 \\ -i & 0 & 0 \end{pmatrix}, \quad \tau_3 = \begin{pmatrix} 0 & -i & 0 \\ i & 0 & 0 \\ 0 & 0 & 0 \end{pmatrix}. \tag{a7}$$

Now let us assume that both $\varepsilon_r$ and $\mu_r$ are constants, for the moment we have

$$\zeta^\rho = \eta^\rho = 0, \quad A^\rho = 0, \quad D_\rho = \partial_\rho = (\partial/u\partial t, \nabla), \tag{a8}$$

and then the Dirac-like equation becomes

$$i\beta^\rho \partial_\rho \psi(x) = 0, \text{ or } i\hbar \frac{\partial}{\partial t}\psi = \hat{H}\psi, \tag{a9}$$

where

$$\hat{H} = -i\hbar u \boldsymbol{\alpha} \cdot \nabla, \quad \boldsymbol{\alpha} = \beta^0 \boldsymbol{\beta} = \begin{pmatrix} 0 & \boldsymbol{\tau} \\ \boldsymbol{\tau} & 0 \end{pmatrix}. \quad (a10)$$

Inside the discontinuous waveguide discussed in the text, as $z < 0$ or $z > L$, for example, one has $\psi = \psi_{TE_{mn}}$ for the $TE_{mn}$ mode, while $\psi = \psi_{TM_{mn}}$ for the $TM_{mn}$ mode, where

$$\psi_{TE_{mn}} = H_{0,mn} \frac{\sqrt{\mu_0 \mu_r}}{\sqrt{2} k_{\perp mn}^2} \begin{pmatrix} -i\frac{\omega}{u}\frac{n\pi}{b}\cos(\frac{m\pi}{a}x)\sin(\frac{n\pi}{b}y) \\ i\frac{\omega}{u}\frac{m\pi}{a}\sin(\frac{m\pi}{a}x)\cos(\frac{n\pi}{b}y) \\ 0 \\ h_{mn}\frac{m\pi}{a}\sin(\frac{m\pi}{a}x)\cos(\frac{n\pi}{b}y) \\ h_{mn}\frac{n\pi}{b}\cos(\frac{m\pi}{a}x)\sin(\frac{n\pi}{b}y) \\ ik_{\perp mn}^2 \cos(\frac{m\pi}{a}x)\cos(\frac{n\pi}{b}y) \end{pmatrix} \exp[-i(\omega t - h_{mn} z)], \quad (a11)$$

$$\psi_{TM_{mn}} = E_{0,mn} \frac{\sqrt{\varepsilon_0 \varepsilon_r}}{\sqrt{2} k_{\perp mn}^2} \begin{pmatrix} ih_{mn}\frac{m\pi}{a}\cos(\frac{m\pi}{a}x)\sin(\frac{n\pi}{b}y) \\ ih_{mn}\frac{n\pi}{b}\sin(\frac{m\pi}{a}x)\cos(\frac{n\pi}{b}y) \\ k_{\perp mn}^2 \sin(\frac{m\pi}{a}x)\sin(\frac{n\pi}{b}y) \\ \frac{\omega}{u}\frac{n\pi}{b}\sin(\frac{m\pi}{a}x)\cos(\frac{n\pi}{b}y) \\ -\frac{\omega}{u}\frac{m\pi}{a}\cos(\frac{m\pi}{a}x)\sin(\frac{n\pi}{b}y) \\ 0 \end{pmatrix} \exp[-i(\omega t - h_{mn} z)]. \quad (a12)$$

It is easy to prove that the solutions given by Eqs. (a11) and (a12) satisfy the Dirac-like equation (a9). In fact, Eqs. (a11) and (a12) are equivalent to Eqs. (1) and (2) in the body of the paper. Let $\psi^\dagger$ denote the Hermitian adjoint of $\psi$, using

$$\int_0^a \cos(\frac{m\pi}{a}x)\cos(\frac{p\pi}{a}x)dx = \int_0^a \sin(\frac{m\pi}{a}x)\sin(\frac{p\pi}{a}x)dx = \frac{a}{2}\delta_{mp}, \quad (a13)$$

$$\int_0^b \cos(\frac{n\pi}{b}y)\cos(\frac{q\pi}{b}y)dy = \int_0^b \sin(\frac{n\pi}{b}y)\sin(\frac{q\pi}{b}y)dy = \frac{b}{2}\delta_{nq}, \quad (a14)$$

one can prove that the TE and TM modes are orthogonal,

$$\int_0^b \int_0^a \psi_{TE_{mn}}^\dagger \psi_{TM_{pq}} \, dxdy = 0. \tag{a15}$$

One of the main advantages of the $(1,0)\oplus(0,1)$ spinor formalism is that it describes the electromagnetic field as a whole, and as the solution of Maxwell equations, it satisfies all the constraints between field components. In contrast, for convenience people usually focus on only one component of electromagnetic waves and study all possible solutions of this single component's equation (e.g., in non-Hermitian photonics). However, the true physical solutions must simultaneously satisfy five other equations, and some of the so-called miraculous results are actually spurious ones.

## APPENDIX B

### Basis functions inside a rectangular waveguide

Eqs. (1) and (2) (or Eqs. (a11) and (a12)) imply that the electromagnetic waves inside the rectangular waveguide can be expanded by the following basis functions:

$$\begin{cases} g_{mn}^{(1)}(x,y) = \cos(\frac{m\pi}{a}x)\sin(\frac{n\pi}{b}y), \ g_{mn}^{(2)}(x,y) = \sin(\frac{m\pi}{a}x)\cos(\frac{n\pi}{b}y) \\ g_{mn}^{(3)}(x,y) = \cos(\frac{m\pi}{a}x)\cos(\frac{n\pi}{b}y), \ g_{mn}^{(4)}(x,y) = \sin(\frac{m\pi}{a}x)\sin(\frac{n\pi}{b}y) \end{cases}. \tag{b1}$$

One can prove that each of them can respectively form an orthonormalized and complete basis. In fact, in consideration of $k_x = \pm m\pi/a$ and $k_y = \pm n\pi/b$, to express that the completeness of $g_{mn}^{(j)}(x,y)$, $j=1,2,3,4$, let us extend the non-negative integers of $m$ and $n$ to $m, n = 0, \pm 1, \pm 2, ...$, such that the transformations of $m \rightarrow -m$ and $n \rightarrow -n$ correspond to those of $k_x \rightarrow -k_x$ and $k_y \rightarrow -k_y$, respectively. Let $0 < x, x' < a$, $0 < y, y' < b$, it is easy to show that:

$$\begin{cases} \sum_{m=-\infty}^{\infty} \cos(\frac{m\pi}{a}x)\cos(\frac{m\pi}{a}x') = \sum_{m=-\infty}^{\infty} \sin(\frac{m\pi}{a}x)\sin(\frac{m\pi}{a}x') = a\delta(x-x') \\ \sum_{n=-\infty}^{\infty} \cos(\frac{n\pi}{b}y)\cos(\frac{n\pi}{b}y') = \sum_{n=-\infty}^{\infty} \sin(\frac{n\pi}{b}y)\sin(\frac{n\pi}{b}y') = b\delta(y-y') \\ \sum_{m=-\infty}^{\infty} \cos(\frac{m\pi}{a}x)\sin(\frac{m\pi}{a}x') = \sum_{n=-\infty}^{\infty} \cos(\frac{n\pi}{b}y)\sin(\frac{n\pi}{b}y') = 0 \end{cases} \quad (b2)$$

and then we have four completeness formulae (note that $[g(x,y)]^{\dagger} = g(x,y)$):

$$\sum_{mn} g_{mn}^{(j)}(x,y) g_{mn}^{(j)}(x',y') = ab\delta(x-x')\delta(y-y'), \quad j=1,2,3,4.$$

More generally, one has

$$\sum_{mn} g_{mn}^{(i)}(x,y) g_{mn}^{(j)}(x',y') = ab\delta(x-x')\delta(y-y')\delta_{ij}, \quad i,j=1,2,3,4. \quad (b3)$$

Likewise, to show that the orthonormality of $g_{mn}^{(j)}(x,y)$, $j=1,2,3,4$, one must take into account the fact that $k_x = \pm m\pi/a$ and $k_y = \pm n\pi/b$. By keeping $m$ and $n$ as non-negative integers, one can extend $x \in [0,a]$ to $x \in [-a,a]$, and $y \in [0,b]$ to $x \in [-b,b]$, in the sense of $a \to -a \Leftrightarrow k_x \to -k_x$, $b \to -b \Leftrightarrow k_y \to -k_y$. Excepting the cases of $m = m' = 0$ and $n = n' = 0$, one has four orthonormality formulae:

$$\begin{cases} \int_{-a}^{a} \cos(\frac{m\pi}{a}x)\cos(\frac{m'\pi}{a}x)dx = \int_{-a}^{a} \sin(\frac{m\pi}{a}x)\sin(\frac{m'\pi}{a}x)dx = a\delta_{mm'} \\ \int_{-b}^{b} \cos(\frac{n\pi}{b}y)\cos(\frac{n'\pi}{b}y)dy = \int_{-b}^{b} \sin(\frac{n\pi}{b}y)\sin(\frac{n'\pi}{b}y)dy = b\delta_{nn'} \\ \int_{-a}^{a} \cos(\frac{m\pi}{a}x)\sin(\frac{m'\pi}{a}x)dx = \int_{-b}^{b} \cos(\frac{n\pi}{b}x)\sin(\frac{n'\pi}{b}x)dx = 0 \end{cases} \quad (b4)$$

and then one has

$$\int_{-a}^{a}\int_{-b}^{b} g_{mn}^{(j)}(x,y) g_{m'n'}^{(j)}(x,y) dxdy = ab\delta_{mm'}\delta_{nn'}, \quad j=1,2,3,4. \quad (b5)$$

More generally, one has

$$\int_{-a}^{a}\int_{-b}^{b} g_{mn}^{(i)}(x,y) g_{m'n'}^{(j)}(x,y) dxdy = ab\delta_{mm'}\delta_{nn'}\delta_{ij}, \quad i,j=1,2,3,4. \quad (b6)$$

In a word, the basis functions of $g_{mn}^{(j)}(x,y)$ ($j=1,2,3,4$), describing the eigenmodes of the waveguide, are respectively orthonormal and complete. Moreover, one has

$$\int_{-\infty}^{\infty} \exp[i(k_z - k_z')z]dz = 2\pi\delta(k_z - k_z'), \quad \int_{-\infty}^{\infty} \exp[ik_z(z-z')]dk_z = 2\pi\delta(z-z'). \quad \text{(b7)}$$

Even so, the expansion form of electromagnetic waves inside a waveguide depends on not only the orthonormality and completeness of basis functions, but also the boundary conditions of the waveguide. Because of the latter, one can expand the electromagnetic waves by means of a subset of basis functions. Therefore, let us come back to the case in which $x \in [0,a]$ and $y \in [0,b]$ for $z<0$ and $z>L$, $x \in [0,s]$ and $y \in [0,d]$ for $0<z<L$, and $m, n = 0,1,2,...$ in the modes $\text{TE}_{mn}$ and $\text{TM}_{mn}$. For example, if we discuss the $\text{TE}_{m0}$ mode, what we need to be concerned about would be the following formulae ($x, x' > 0$, $y, y' > 0$, $m, m' = 1, 2, ...$):

$$\sum_{m=1}^{\infty} \cos(\frac{m\pi}{a}x)\cos(\frac{m\pi}{a}x') = \frac{1}{2}[a\delta(x-x')-1], \quad \text{(b8)}$$

$$\sum_{m=1}^{\infty} \sin(\frac{m\pi}{a}x)\sin(\frac{m\pi}{a}x') = \frac{a}{2}\delta(x-x'), \quad \text{(b9)}$$

$$\int_0^a \cos(\frac{m\pi}{a}x)\cos(\frac{m'\pi}{a}x)dx = \int_0^a \sin(\frac{m\pi}{a}x)\sin(\frac{m'\pi}{a}x)dx = \frac{a}{2}\delta_{mm'}. \quad \text{(b10)}$$

---